\documentclass[twocolumn,pre,amsmath,showpacs,amssymb,amsfonts,10pt]{revtex4-1}

\usepackage{graphicx}
\usepackage{subfigure}
\usepackage[english]{babel}

\begin{document}

\title{
Ising-XY Transition in Three-dimensional Frustrated Antiferromagnets with Collinear Ordering
}
\author{A.O.~Sorokin$^{1,2}$}
\email{aosorokin@gmail.com}
\affiliation{$^1$Department of Physics, St.\ Petersburg State University, 198504 St.\ Petersburg, Russia}
\affiliation{$^2$Petersburg Nuclear Physics Institute, NRC Kurchatov Institute, 188300 St.\ Petersburg, Russia}

\date{\today}

\begin{abstract}
 Using Monte Carlo simulations, we study the critical behavior of two models of frustrated XY antiferromagnets with a collinear spin ordering and with an additional twofold degeneracy of the ground state. We consider a classic antiferromagnet on a body-centered cubic lattice with an additional antiferromagnetic exchange interaction between next-nearest spins, and a ferromagnet with an extra antiferromagnetic intralayer exchange. In both models, a single first-order transition on the discrete and continuous order parameters is found. Observed critical pseudo-exponents are in agreement with exponents of XY magnets with a planar spin ordering like a stacked-triangular antiferromagnet and helimagnets belonging to the same symmetry class.
\end{abstract}

\pacs{64.60.De, 75.40.Cx, 05.10.Ln, 75.10.Hk}

\maketitle

\section{Introduction}

Critical phenomena in frustrated magnetic systems are very attractive due to realization in them a plurality of symmetry breaking scenarios \cite{Loison04}. During the last several decades, it have been discussed a possibility that the new universality class different from the class of the usual $O(N)$ model is realized in magnets with planar spin  ordering like a stacked-triangular antiferromagnet (STA) and helimagnets \cite{Kawamura88}. In such systems, the $O(N)/O(N-2)$ symmetry is broken, whereas in the $O(N)$ model the broken symmetry is $O(N)/O(N-1)$. In the case of XY spins ($N=2$) and a planar ordering, the broken $O(2)=\mathbb{Z}_2\otimes SO(2)$ symmetry relates to global spin rotations and inversions. In three-dimensional systems belonging to the later symmetry class, one observe a single fluctuation-induced first-order transition describing by the Ginzburg-Landau functional (the $O(N)\otimes O(2)$-model)
\begin{eqnarray}
    F=&&\int d^3x\left((\partial_\mu\mathbf{\phi}_1)^2+(\partial_\mu\mathbf{\phi}_2)^2+r(\mathbf{\phi}_1^2+\mathbf{\phi}_2^2)+
    \right.\nonumber\\
    &&\left.
    u\left(\mathbf{\phi}_1^2+\mathbf{\phi}_2^2\right)^2+
    2w\left((\mathbf{\phi}_1\mathbf{\phi}_2)^2-\mathbf{\phi}_1^2\mathbf{\phi}_2^2\right)\right),
    \label{GLW-model}
\end{eqnarray}
where $\mathbf{\phi}_{1,2}$ are two-component vectors, $u,\,w>0$ and $u>w$. This transition is close to a second-order transition, therefore at the critical point, one may observe (pseudo) scaling behavior and universality. Such an imitation of a second-order transition behavior is typical for weak first-order transitions, e.g. as in the two-dimensional 5-state Potts model \cite{Landau89}.

In terms of the renormalization group (RG) \cite{Zumbach93, Delamotte00}, the fixed point in the model \eqref{GLW-model}, being stable for large $N$, becomes complex-valued for the case $N=2$, with the small imaginary part. On the real part of the RG-diagram in a vicinity of this fixed point, the region exists where the RG-flow is slow. It leads to imitation of the scaling behavior. This region is attractive in sense that trajectories starting from a sufficiently wide area of initial point cross it. The studies of the model \eqref{GLW-model} at $N=2$ \cite{Zumbach93, Delamotte00} have shown that the region of the significant RG-flow slowdown is quite large, so one may observe a wide scatter of critical pseudo-exponents values for different lattice models. Furthermore, some models from considered symmetry class have a distinct first order of a transition (see, e.g. \cite{Loison98}).

In this work, we consider two models of frustrated XY antiferromagnets with a {\it collinear} spin ordering and with an additional broken $\mathbb{Z}_2$ subgroup of a lattice symmetry, so the full broken symmetry is also $\mathbb{Z}_2\otimes SO(2)$. The first model is an antiferromagnet on a body-centered cubic lattice with an extra antiferromagnetic exchange interaction between next-nearest spins. This model is equivalent to two interacting antiferromagnetic sublattices. Due to competing of exchanges, the additional twofold degeneracy of the ground state appears \cite{Shender82}. Note that models of classic XY antiferromagnets with competing exchanges have been investigated for other lattices, e.g. a simple cubic \cite{Pinnetes}, a face-centered cubic \cite{Diep89}, a hexagonal closed-packed \cite{Diep92}, and a stacked-triangular \cite{Loison93} lattices. For these models, an extra degeneracy of a ground state is threefold, and a transition of distinct first order occurs with breaking of the $\mathbb{Z}_3\otimes SO(2)$ symmetry.

The second model is so-called stacked-J$_1$-J$_2$ model of a ferromagnet on a simple cubic lattice with an additional intralayer exchange, where we have a twofold degeneracy of the ground state \cite{Henley}. This model is intensively investigated in two dimensions in a vicinity of the quantum critical point. For classical spins, this model has been studied numerically in \cite{Loison00}, where it has been found that transitions on the continuous and discrete order parameters occur at different temperatures. Generally, in two-dimensional models from the same symmetry class, transitions on both order parameters occur separately in temperature or coalesce into a single transition of first order \cite{Korshunov}.

The considered lattice models belong to the universality class with the broken $\mathbb{Z}_2\otimes SO(2)$ symmetry being discussed above, but they are described by another functional (the $\mathbb{Z}_2\otimes O(N)$-model) \cite{SorokinFut}
\begin{eqnarray}
    F=&&\int d^3x\left((\partial_\mu\mathbf{\phi}_1)^2+(\partial_\mu\mathbf{\phi}_2)^2+r(\mathbf{\phi}_1^2+\mathbf{\phi}_2^2)+
    \right.\nonumber\\
    &&\left.
    u\left(\mathbf{\phi}_1^4+\mathbf{\phi}_2^4\right)+
    2w(\mathbf{\phi}_1\mathbf{\phi}_2)^2+2v\mathbf{\phi}_1^2\mathbf{\phi}_2^2\right),
    \label{GLW-model-2}
\end{eqnarray}
with $u>0$, $w<0$ and $u+v+w>0$. For the case $N=2$, the models \eqref{GLW-model} and \eqref{GLW-model-2} are equivalent in the mean-field approximation, but critical fluctuation in them are different. Nevertheless, we find a transition of a weak first order in both lattice models. Moreover, we observe the critical pseudo-exponents, which are close to the exponents of STA and helimagnets. The comparing of results is shown in table \ref{tab}.

\section{Models and methods}

\begin{figure}[t]
\includegraphics[scale=0.35]{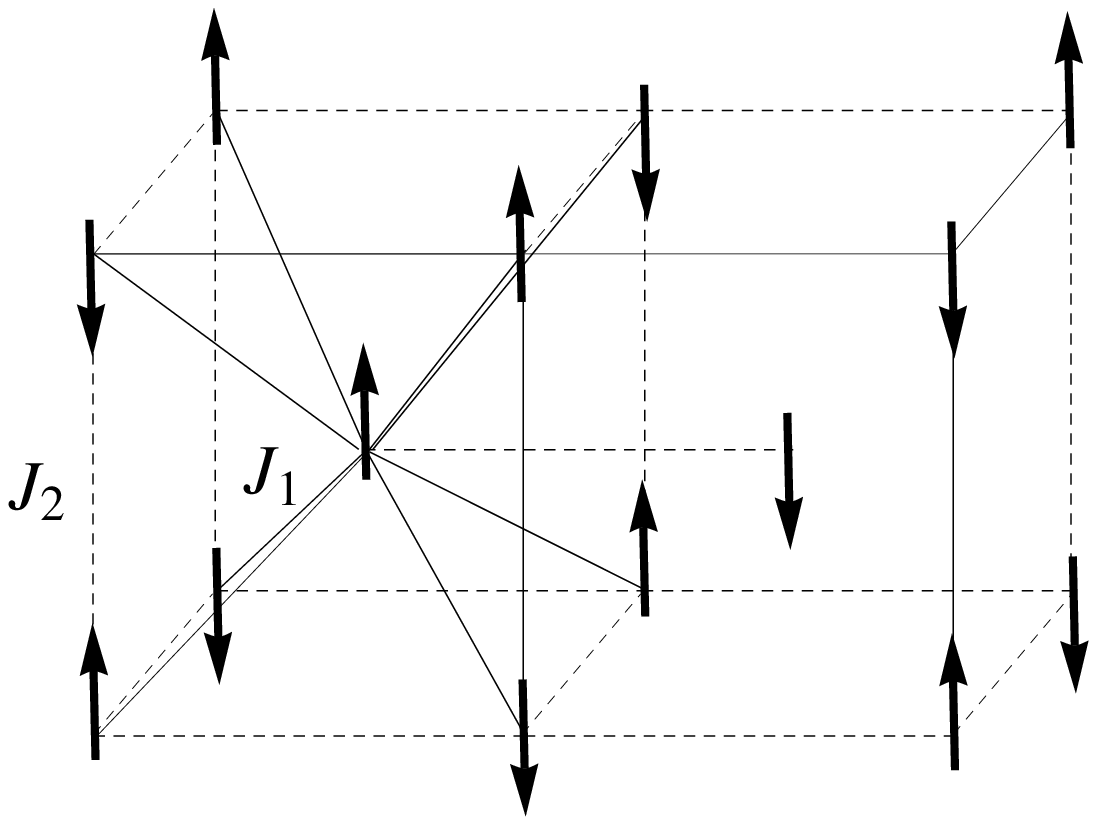}
\includegraphics[scale=0.35]{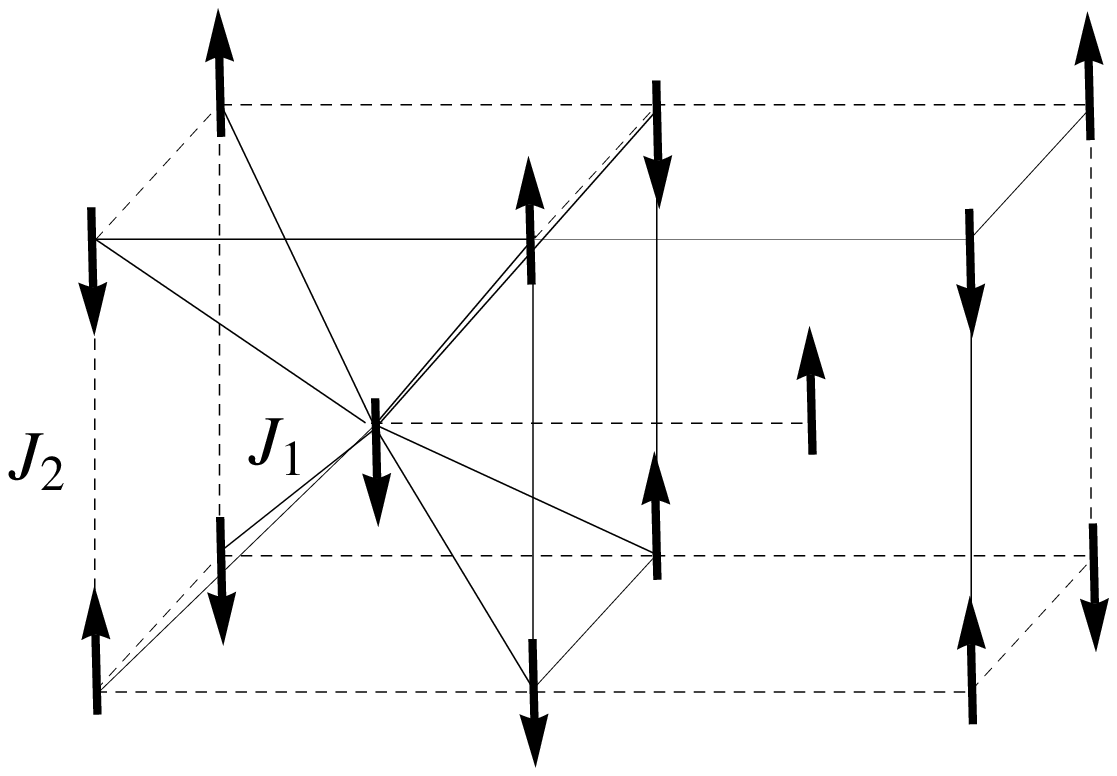}
\caption{\label{fig1} Exchange interactions in the ABC model. We show two configurations with opposite chirality irreducible to each other by global spin rotations and inversions.}
\end{figure}%
The model of an antiferromagnet on a body-centered cubic lattice (for brevity, the ABC-model) is described by the Hamiltonian
\begin{equation}
    H=J_1\sum_{ij}\mathbf{S}_i\mathbf{S}_j+J_2\sum_{kl}\mathbf{S}_k\mathbf{S}_l,
    \label{model2}
\end{equation}
where the sum $ij$ runs over pairs of nearest spins of a lattice, and the sum $kl$ runs over nest-nearest spins (fig. \ref{fig1}). A spin $\mathbf{S}$ is a classic 2-component vector, $J_1,\,J_2>0$. When $J_2<2J_1/3$, the ground state is two ferromagnetic sublattices interacting antiferromagnetically, and the frustration does not appear. But when $J_2>2J_1/3$, sublattices become antiferromagnetic. Two non-equivalent ground states are present in fig. \ref{fig1}.

\begin{figure}[t]
\includegraphics[scale=0.35]{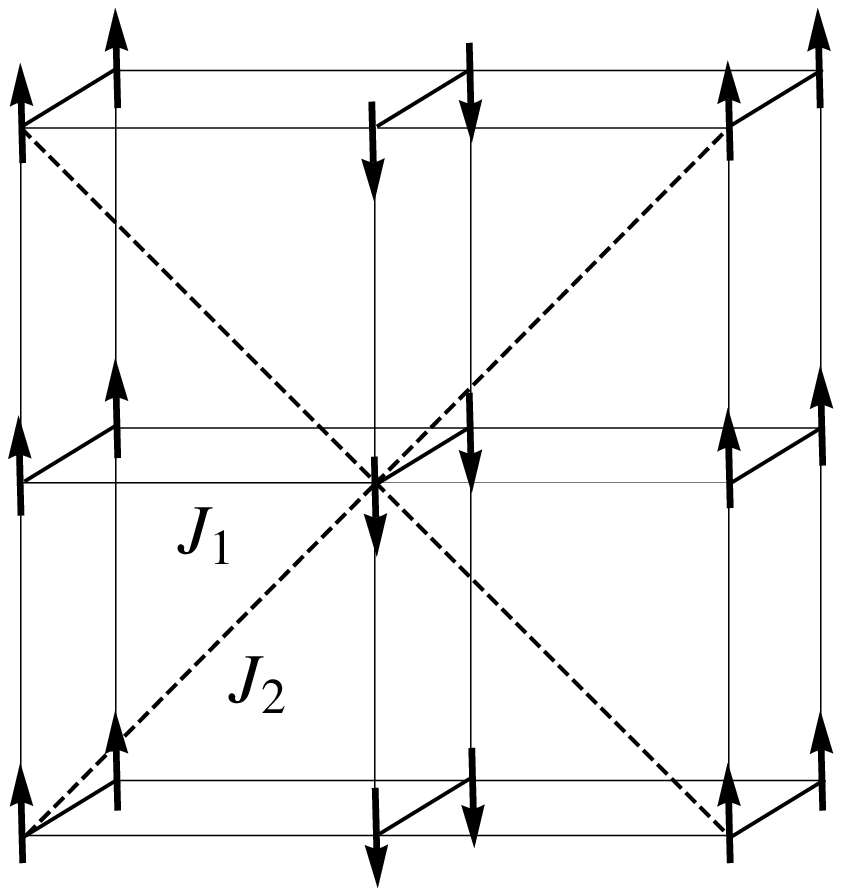}
\includegraphics[scale=0.35]{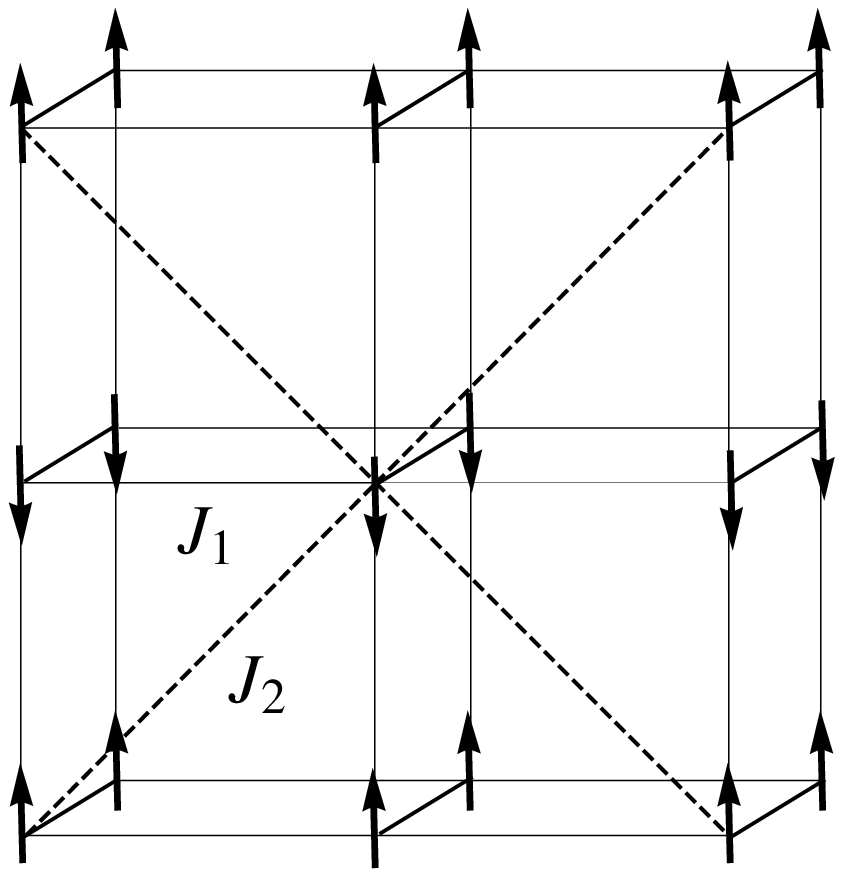}
\caption{\label{fig2} Exchange interactions in the SJJ model. We show two configurations with opposite chirality irreducible to each other by global spin rotations and inversions.}
\end{figure}
Another model, called as the SJJ model, is
\begin{equation}
    H=-J_1\sum_{ij}\mathbf{S}_i\mathbf{S}_j+J_2\sum_{kl}\mathbf{S}_k\mathbf{S}_l,
    \label{model1}
\end{equation}
where the sum $ij$ runs over nearest spin pairs of simple cubic lattice, and the sum $kl$ enumerates next-nearest spin pairs in layers (fig. \ref{fig2}). When $J_2<J_1/2$, the ground state is a ferromagnetic ordering. When $J_2>J_1/2$, two type of configurations with wave-vectors $\mathbf{q}=(\pi,0,0)$ and $\mathbf{q}=(0,\pi,0)$ correspond to the ground state. This model is equivalent to two simple tetragonal antiferromagnetic sublattices with lattice vectors $(1,1,0)$, $(1,-1,0)$ and $(0,0,1)$, embedded into each other and shifted relative to one another with the vector $(1,0,0)$. Wherein the exchange $J_2$ is related to an internal interaction of sublattice, and $J_1$ is an interaction between sublattices.

The proposed models \eqref{model1} and \eqref{model2} are studied by Monte Carlo simulations using the over-relaxed algorithm \cite{algorithm}. To define the order of a transition, we use the histogram analysis method. Thermalization is performed within $2\cdot10^5$ Monte Carlo steps per spin, and calculation of averages, within $3.4\cdot10^6$ steps. We use periodic boundary conditions and consider lattices with sizes $16\leq L\leq 100$. The values of exchanges are chosen as $J_1=J_2=1$.

The magnetic order parameter in both models is define using four sublattices
\begin{equation}
    \mathbf{m}_i=\frac{4}{L^3}\sum_{\mathbf{x}_i}\mathbf{S}_{\mathbf{x}_i},
    \quad \bar m=\sqrt{\frac1{4}\sum_{i=1}^{4}
    \langle\mathbf{m}_i^2\rangle},
    \label{3D-m}
\end{equation}
where $\mathbf{x}_i$ runs over cites of the $i$-th sublattice, $L^3$ is the volume of the system (a number of spins in the ABC model is equal to  $2L^3$). The chirality in the SJJ model is the following
$$
k=\frac1{4L^3}\sum_\mathbf{x}\left(\mathbf{S}_\mathbf{x}-\mathbf{S}_{\mathbf{x}+\mathbf{e}_1+\mathbf{e}_2}\right)
                    \left(\mathbf{S}_{\mathbf{x}+\mathbf{e}_1}-\mathbf{S}_{\mathbf{x}+\mathbf{e}_2}\right),
$$
\begin{equation}
    \bar k=\langle |k|\rangle,
\end{equation}
with $\mathbf{e}_\mu$ is a unit vector along corresponding direction of a lattice. In the ABC model, the chiral order parameter is defined as
\begin{eqnarray}
    k=\frac1{8L^3}\sum_\mathbf{x}\mathbf{S}_{\mathbf{x}+\mathbf{a}}\left(\mathbf{S}_\mathbf{x}-\mathbf{S}_{\mathbf{x}+\mathbf{e}_1}-\mathbf{S}_{\mathbf{x}+\mathbf{e}_2}
    -\mathbf{S}_{\mathbf{x}+\mathbf{e}_3}+\right.\nonumber\\
    \left.+\ \mathbf{S}_{\mathbf{x}+\mathbf{e}_1+\mathbf{e}_2}+\mathbf{S}_{\mathbf{x}+\mathbf{e}_1+\mathbf{e}_3}
    +\mathbf{S}_{\mathbf{x}+\mathbf{e}_2+\mathbf{e}_3}-\mathbf{S}_{\mathbf{x}+\mathbf{e}_1+\mathbf{e}_2+\mathbf{e}_3}\right),\nonumber\\
\end{eqnarray}
with $\mathbf{a}=(\frac12,\frac12,\frac12)$ is the shift vector of antiferromagnetic sublattices.
\begin{figure}[t]
\includegraphics[scale=0.4]{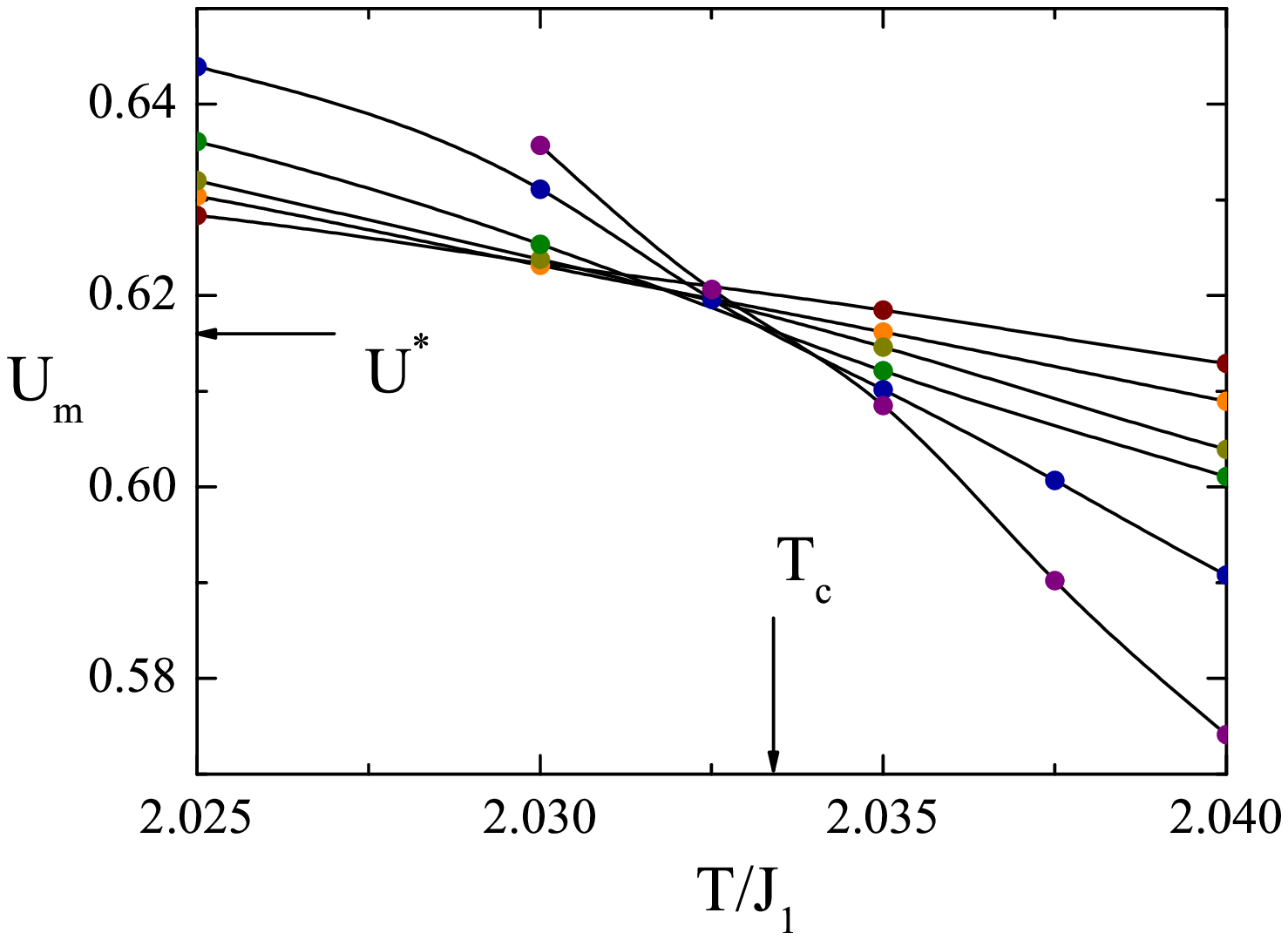}
\includegraphics[scale=0.4]{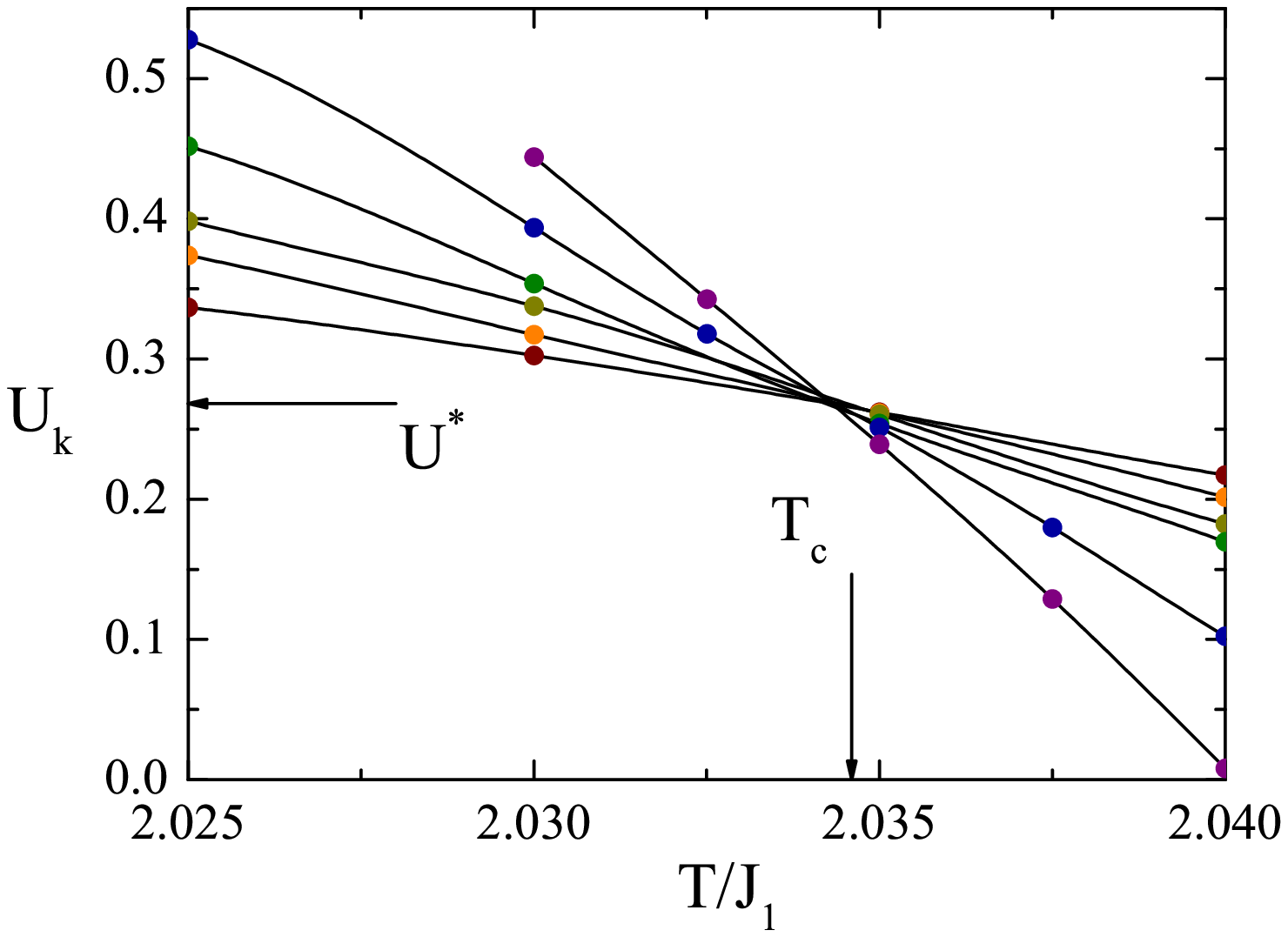}
\caption{\label{fig3}Estimation of the transition temperature in the SJJ model.}
\end{figure}

To estimate the transition temperature on one of the order parameters $p=\bar m, \bar k$, as well as the precision of the estimation procedure, we use the Binder cumulant
crossing method \cite{Binder}
\begin{equation}
    U_p=1-\frac{\langle p^4 \rangle}{3\langle p^2 \rangle^2}.
\end{equation}
Critical exponent $\nu$ is estimated using the following cumulant \cite{Ferren}
\begin{equation}
V_p=\frac{\partial}{\partial(1/T)}\ln
\langle p^2\rangle=L^2\left(\frac{\left<p^2E\right>}{\left<p^2\right>}-\langle E\rangle\right),
\label{Vp}
\end{equation}
and besides, we duplicate the estimation procedure using all order parameters. The scaling relations are
\begin{equation}
    \max\left(V_p^{(n)}\right)\sim L^{\frac1\nu},\quad
    \left.\bar p\right|_{T=T_c}\sim L^{-\frac\beta\nu},\quad
    \left.\chi_p\right|_{T=T_c}\sim L^{\frac\gamma\nu},
\end{equation}
with $\chi_p$ is susceptibility corresponding to the order parameter $p$ \cite{Binder}
\begin{eqnarray}
    \chi_p=\frac{L^d}{T}\Bigl(\left<p^2\right>-\langle
    |p|\rangle^2\Bigr),\quad T<T_c;\nonumber\\
    \chi_p=\frac{L^d}{T}\left<p^2\right>,\quad T\geq T_c.
\end{eqnarray}
\begin{figure}[t]
\includegraphics[scale=0.4]{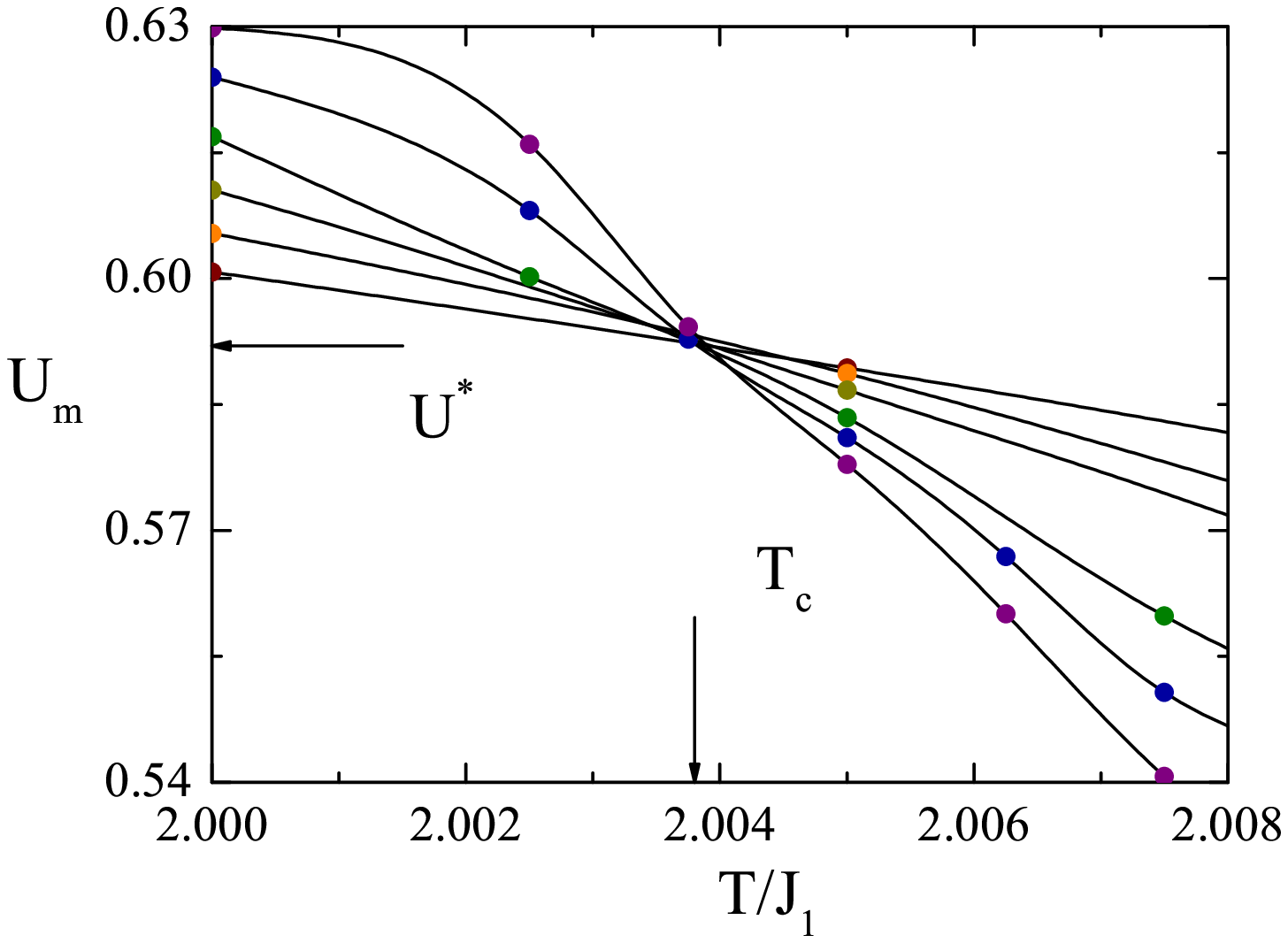}
\includegraphics[scale=0.4]{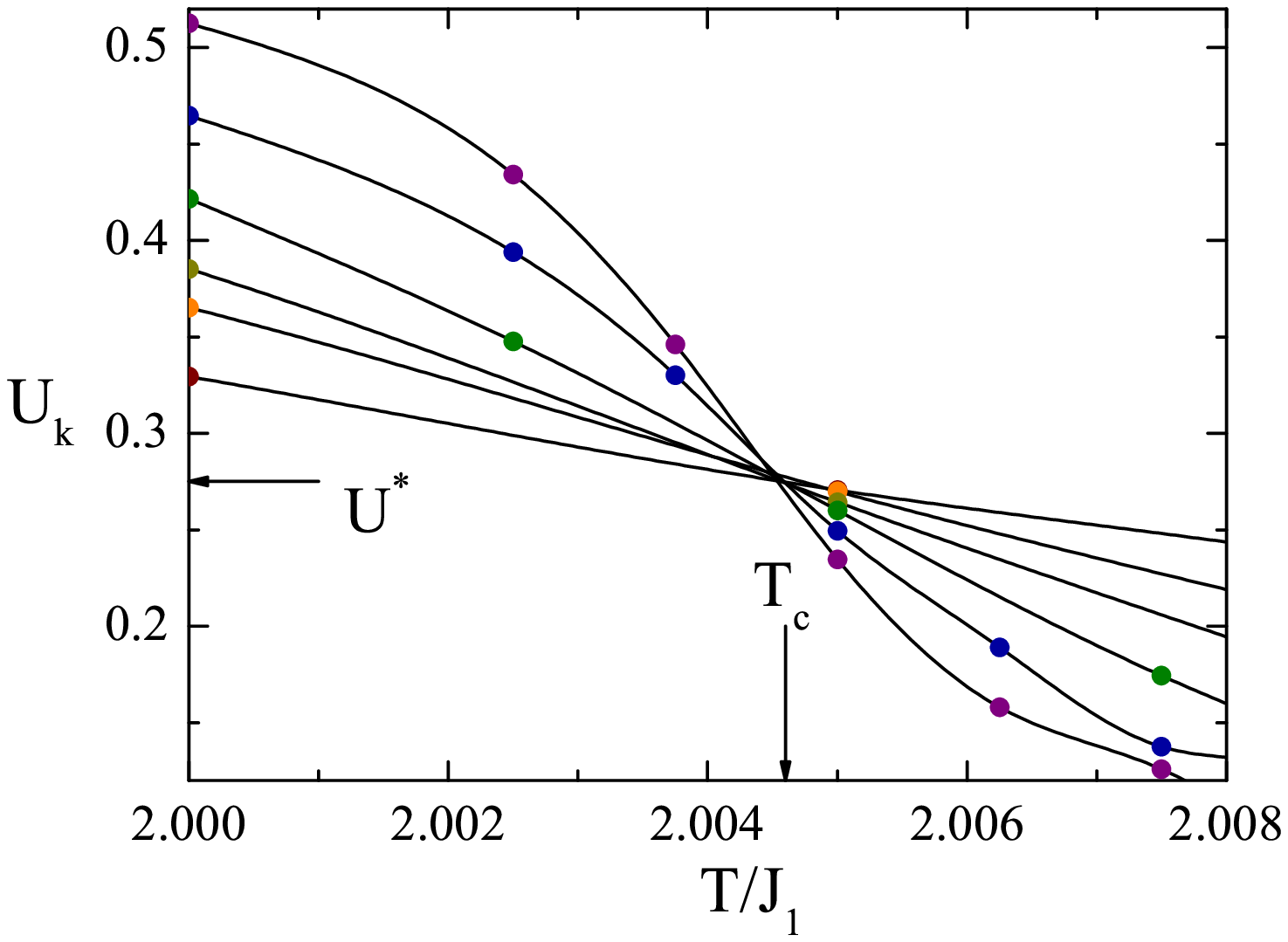}
\caption{\label{fig4}Estimation of the transition temperature in the ABC model.}
\end{figure}%

\section{Results}

In the SJJ model with $N=2$, the transition temperatures on the magnetic and chiral order parameters are estimated as (fig. \ref{fig3}):
\begin{equation}
    T_c^{(m)}/J_1=2.0334(12),\quad
    T_c^{(k)}/J_1=2.0346(10),
\end{equation}
with the values of The Binder cumulant
\begin{equation}
    U_m^*=0.616(5),\quad
    U_k^*=0.268(8).
\end{equation}
Thus, within the accuracy of the results, we are dealing with a single transition. Similarly, for the ABC model we find (fig. \ref{fig4})
\begin{equation}
    T_c^{(m)}/J_1=2.0038(7),\quad
    T_c^{(k)}/J_1=2.0046(9),
\end{equation}
\begin{equation}
    U_m^*=0.592(4),\quad
    U_k^*=0.275(7).
\end{equation}
Note that the value of the Binder cumulant at a critical point is expected to be universal depending on boundary conditions while a transition is of second order. However, for the three-dimensional XY helimagnet, the values $U_m^*=0.623(7)$ and $U_k^*=0.39(3)$ have been found \cite{Sorokin14}, which contradict to the present results. It is more unexpectedly that the critical exponents describing the pseudo-scaling behavior of the considered models are in agreement with the known results of STA and helimagnets. Table \ref{tab} shows the exponent estimations and their comparison with critical indices of other symmetry classes.
\begin{table*}[th]
\caption{\label{tab}Comparison of the (pseudo) critical exponents obtained in the present work (marked as $[\ast]$) with known exponents of other universality classes. Notations: Ising --- the Ising model, Ferro --- ferromagnetic or the $O(N)$ model, Helix --- helimagnets, STA --- antiferromagnet on a stacked-triangular lattice, FSS --- typical exponents for a first-order transition in the finite-size-scaling theory.}
\begin{ruledtabular}
\begin{tabular}{ccc|cccccc}
Class $G/H$     & Model & Ref.    &$\nu$   & $\nu_k$ & $\beta$ & $\beta_k/2$    & $\gamma$ &$\gamma_k+\beta_k$\\
\hline
$\mathbb{Z}_2$   &Ising & \cite{Vicari02}      &0.630   &         & 0.327   &                & 1.236    & \\
$SO(2)$          &Ferro & \cite{Vicari02}      &0.671   &         & 0.348   &                & 1.317    & \\
$SO(3)/SO(2)$    &Ferro & \cite{Vicari02}      &0.706   &         &0.365    &                & 1.388    &\\
$SO(4)/SO(3)$    &Ferro & \cite{Vicari02}      &0.75    &         &0.39     &                &1.47      &\\
Tricritical   & Mean-field &                   &0.5     & 0.5     &0.25     & 0.25           &1.00      &1.00\\
First order   & FSS    &                       &0.33    & 0.33    &0        & 0              &1.00      &1.00\\
\hline
$\mathbb{Z}_2\otimes SO(2)$& Helix & \cite{Sorokin14}  &0.55     & 0.56     & 0.25    & 0.21           & 1.16     & 1.29\\
$\mathbb{Z}_2\otimes SO(2)$& STA   & \cite{Kawamura92} &0.54     & 0.55     & 0.25    & 0.23           & 1.13     & 1.22\\
$\mathbb{Z}_2\otimes SO(2)$& SJJ & $[\ast]$    &0.565(8) & 0.572(10)& 0.260(6)& 0.251(8)       & 1.18(4)  & 1.22(5)\\
$\mathbb{Z}_2\otimes SO(2)$& ABC & $[\ast]$            &0.568(10)& 0.571(9) & 0.262(7)& 0.258(10)      & 1.18(5)  & 1.20(8)\\

\end{tabular}
\end{ruledtabular}
\end{table*}%

\begin{figure}[t]
\includegraphics[scale=0.38]{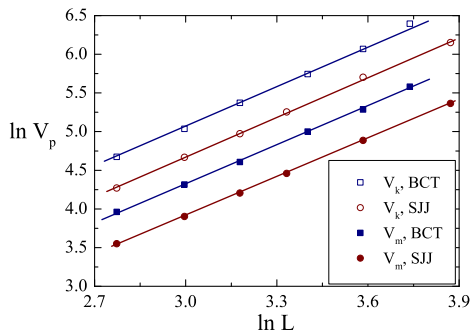}
\caption{\label{fig5}Estimation of the exponents $\nu_p$ using the scaling of the cumulant \eqref{Vp} for both models.}
\end{figure}
\begin{figure}[t]
\includegraphics[scale=0.4]{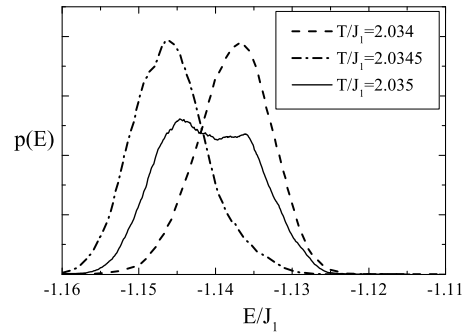}
\caption{\label{fig6}Internal energy distribution for $L=90$ near the critical temperature in the SJJ model.}
\end{figure}

%
We find that the phase transition is of first order. To observe a jump of the internal energy typical for first-order transitions, we consider large lattices with $L=80,\ 90$ è $100$. Similar sizes of a lattice have been used to determine the transition order in STA \cite{Diep08} and helimagnets \cite{Sorokin14}. The double-peak structure of energy distribution at the critical point indicates the presence of internal heat of a transition, it is shown in fig. \ref{fig6}.

\section{Discussion}

The $\mathbb{Z}_2\otimes SO(2)$ symmetry, broken in a ordered phase of XY magnetic systems with a planar spin ordering, is a symmetry group acting in the spin space and related to global spin rotations and inversions. For spins with an arbitrary number of vector components, the $O(N)$ symmetry is broken down to $O(N-2)$ subgroup below a critical temperature. For the systems considered in the present work, this symmetry is broken down to $O(N-1)$ subgroup merely, and the additional $\mathbb{Z}_2$ factor responds to a lattice symmetry breaking. The class $\mathbb{Z}_2\otimes O(N)/O(N-1)$ containing these lattice models and the class $O(N)/O(N-2)$ of magnets with a planar ordering become equivalent just for $N=2$. Nonetheless, even for the case $N=2$, the critical behavior in systems from these classes describes generally by non-equivalent models \eqref{GLW-model} and \eqref{GLW-model-2}. Though both models exhibit a similar critical behavior, namely that a transition is of weak first order.

The model \eqref{GLW-model-2} includes the model \eqref{GLW-model} with $w=u-v>0$ as a particular case. The region of the RG-flow slowdown, explaining the pseudo-scaling behavior in canted magnetic systems, resides in the sector $w>0$. At the same time, the $\mathbb{Z}_2\otimes O(N)/O(N-1)$ symmetry breaking scenario corresponds to the sector $w<0$, where is no fixed point for $N\geq2$, even complex-valued with $\mathrm{Re}\, w<0$ \cite{SorokinFut}. Therefore, a search of a RG-flow slowdown region in the sector $w<0$ must be performed independently on the known results \cite{Zumbach93, Delamotte00}.

The investigations of the model \eqref{GLW-model} with $N=2$ using non-perturbative RG \cite{Zumbach93, Delamotte00} have shown that the region of the RG-flow slowdown is wide enough and does not have a distinct center associated with a local minimum of the RG-flow, in contradistinction to the case $N=3$. Moreover, the region includes a vicinity of the Heisenberg fixed point with $w=0$ responding to the $O(4)$ model. It explains a variation in pseudo-exponent values observed in numerical studies and experiments (see \cite{Delamotte04} for a review). The exponents may take values close to the Kawamura's critical indices for STA \cite{Kawamura92} as well as close to the indices of the $O(4)$ model. We expect that the region of the slow RG-flow continues to the sector $w<0$ and probably encompasses not only a vicinity of the Heisenberg point, but also a vicinity of the decoupled fixed point $w=v=0$ responding to two non-interacting $O(2)$ models. (Fixed points of the model \eqref{GLW-model-2} have been found in \cite{SorokinFut}, and the case $w=0$ has been studied extensively in \cite{Pelissetto03}.) Such a wide region may contain trajectories, where the imitated scaling behavior is described by exponents close to the Kawamura's results, but without the pseudo-universality.

When $w$ is sufficiently small, it is possible to explain the observed pseudo-universality by another way, associated with a tricritical behavior. Such a possibility has been discussed for STA in \cite{plumer94}. In table \ref{tab}, we shows the mean-field values of the tricritical exponents, which are close to the exponents of systems with the breaking $\mathbb{Z}_2\otimes SO(2)$ symmetry. While the term $w(\phi_1\phi_2)^2$ is small, the scaling is determined by terms like $w_6(\phi_1\phi_2)^3$. The later terms are inessential at a critical point, but they can lead to crossover between a tricritical and critical behaviors, discernible in simulations on finite-size lattices or at some distance on temperature from a critical point. Note that such a crossover have been observed in a similar model in two dimensions \cite{Fujimoto}.

One can consider one more possibility associated with an approximation of first-order transition singularities in the finite-size-scaling theory. In table \ref{tab}, we shows also typical "exponents"\ for a first-order transition, but they are disagreement with the exponents of the considered models. Apparently, such an explanation is not applicable to the weak first-order or "almost second-order"\ transitions.

\begin{acknowledgments}
The author acknowledges Saint Petersburg State University for the research grant 11.50.2514.2013.
This work was also supported by the RFBR grant No 14-02-31448.
\end{acknowledgments}

\end{document}